\documentclass[final,5p,times,twocolumn]{elsarticle}

\usepackage[T1]{fontenc}
\usepackage[utf8]{inputenc}

\usepackage{amsmath}
\usepackage{amsfonts}
\usepackage{amssymb}
\usepackage{amsxtra}
\usepackage{array}
\usepackage{color}
\usepackage{dcolumn}
\usepackage{graphicx}
\usepackage{hepunits}
\usepackage{xspace}
\usepackage{bm}

\definecolor{purple}{rgb}{0.5,0,0.5}
\definecolor{blue}{rgb}{0.0,0,0.9}
\definecolor{prdblue}{rgb}{0.133,0.118,0.498}
\usepackage[colorlinks=true, pdfstartview=FitV, linkcolor=prdblue, citecolor= prdblue, urlcolor=prdblue]{hyperref}

\usepackage[mathscr,scaled=1.15]{urwchancal}
\DeclareFontFamily{OT1}{pzc}{}
\DeclareFontShape{OT1}{pzc}{m}{it}
{<-> s * [1.15] pzcmi7t}{}
\DeclareMathAlphabet{\mathpzc}{OT1}{pzc}{m}{it}

\biboptions{sort&compress}

\journal{Physics Letters B}

\makeatletter

\setbox0\hbox{$\xdef\scriptratio{\strip@pt\dimexpr
    \numexpr(\sf@size*65536)/\f@size sp}$}

\newcommand{\scriptveryshortarrow}[1][3pt]{{%
    \hbox{\rule[\scriptratio\dimexpr\fontdimen22\textfont2-.2pt\relax]
               {\scriptratio\dimexpr#1\relax}{\scriptratio\dimexpr.4pt\relax}}%
   \mkern-4mu\hbox{\let\f@size\sf@size\usefont{U}{lasy}{m}{n}\symbol{41}}}}

\makeatother

\begin{document}

\begin{frontmatter}

\title{Pion structure from its light-front wave function}

\author[UHe]{Kh\'epani Raya%
    $^{\href{https://orcid.org/0000-0001-8225-5821}{\textcolor[rgb]{0.00,1.00,0.00}{\sf ID}},}$}

\author[UHe]{Zhen-Ni Xu%
       $^{\href{https://orcid.org/0000-0002-1651-5717}{\textcolor[rgb]{0.00,1.00,0.00}{\sf ID}},}$}

\author[UHe,UPO]{Zhao-Qian Yao%
       $^{\href{https://orcid.org/0000-0002-1651-5717}{\textcolor[rgb]{0.00,1.00,0.00}{\sf ID}},}$}

\author[UHe]{Jos\'e Rodr\'{\i}guez-Quintero%
       $^{\href{https://orcid.org/0000-0002-1651-5717}{\textcolor[rgb]{0.00,1.00,0.00}{\sf ID}},}$}


\address[UHe]{
Dpto.\ Ciencias Integradas, Centro de Estudios Avanzados en Fis., Mat. y Comp., Fac.\ Ciencias Experimentales, Universidad de Huelva, Huelva 21071, Spain}
\address[UPO]{Helmholtz-Zentrum Dresden-Rossendorf, Bautzner Landstra{\ss}e 400, D-01328 Dresden, Germany\\[1ex]
%
\href{mailto:khepani.raya@dci.uhu.es}{khepani.raya@dci.uhu.es} (Kh\'epani Raya);
\href{mailto:jose.rodriguez@dfaie.uhu.es}{jose.rodriguez@dfaie.uhu.es} (Jos\'e Rodr{\'{\i}}guez-Quintero)
%
}

\begin{abstract}
Understanding the structural properties of the pion is essential for elucidating the mechanisms of mass generation within the Standard Model and their role in the emergence and properties of the hadronic matter. Light-front wave functions encode extensive information about the internal structure of these systems and provide the link to measurable quantities such as generalized parton distributions and transverse-momentum-dependent distributions. Guided by recent progress in continuum Schwinger methods, we derive well-founded and practical representations of these quantities, enabling the exploration of several facets of the pion structure, including distribution amplitudes and distribution functions, elastic and gravitational form factors, and the associated momentum and spatial distributions. The results presented here are consistent with expectations and can be tested at modern experimental facilities, including the new generation of electron–ion colliders.
\end{abstract}


\end{frontmatter}


\section{Introduction}
%
The proton has long played a central role in our understanding of the strong interactions described by quantum chromodynamics (QCD),\,\cite{Hofstadter:1956qs,Breidenbach:1969kd}. However, a growing body of evidence now indicates that unraveling the structure of the pion is of comparable importance, e.g. Refs.~\cite{Aguilar:2019teb,Roberts:2021nhw}. Owing to its role as an effective mediator of the nuclear force, as the lightest bound state in QCD, and as the Nambu–Goldstone (NG) boson associated with dynamical chiral symmetry breaking (DCSB), the pion is uniquely suited to probe the emergent facets of QCD\,\cite{Raya:2024ejx}. These include quark–gluon confinement, which gives rise to observable degrees of freedom (hadrons), and the emergence of hadron masses (EHM), responsible for nearly all the mass of visible matter\,\cite{Roberts:2020hiw}.

Generalized parton distributions (GPDs) offer a comprehensive picture of the internal dynamics of hadrons,\,\cite{Mezrag:2022pqk}, granting access to their electromagnetic and mechanical structure,\,\cite{Polyakov:2018zvc}\, and enabling spatial imaging,\,\cite{Diehl:2002he,Burkardt:2000za}. At a deeper level, pion GPDs can be connected to the corresponding light-front wave function (LFWF),\,\cite{Diehl:2000xz,Burkardt:2002hr}, which in turn raises a bridge to complementary information through transverse-momentum-dependent (TMD) distributions\,\cite{Pasquini:2014ppa}. The patterns encoded in GPDs and TMDs, and related observables, are closely tied to the emergent facets of QCD, particularly EHM\,\cite{Zhang:2021mtn,Raya:2021zrz,Raya:2022eqa}. Thus, these objects offer a comprehensive and multidimensional landscape of the pion’s internal structure.

All these quantities are expected to be probed with unprecedented precision at the future Electron–Ion Collider, as well as at other modern experimental facilities\,\cite{Accardi:2012qut,Arrington:2021biu,Anderle:2021wcy,Quintans:2022utc,Accardi:2023chb}. It is therefore timely to develop robust theoretical predictions in anticipation of forthcoming empirical data. Guided by recent progress in continuum Schwinger methods (CSMs), e.g.\,\cite{Zhang:2021mtn,Raya:2021zrz,Raya:2022eqa}, in this work we explore a systematic construction of the LFWF for pseudoscalar mesons, with particular emphasis on the pion. Many aspects of the pion’s structural properties revealed through GPDs, TMDs, and LFWFs, are subsequently discussed.

The manuscript is organized as follows. Section~\ref{sec:LFWF} introduces relevant definitions and concepts for LFWFs. In Sec.~\ref{sec:PTIR}, we present two complementary approaches for deriving the corresponding LFWFs within a framework connected to CSMs. GPDs and the associated form factors are introduced in Sec.~\ref{sec:GPD}, while different aspects of transverse structure (in both momentum and position space) are discussed in Sec.~\ref{sec:Transverse}. Finally, conclusions and scope are presented in Sec.~\ref{sec:conclusions}.

\section{Light-front Wave Functions}
\label{sec:LFWF}

Let us consider a pseudoscalar meson $\textbf{P}=q\bar{h}$, with mass $m_\textbf{P}$. The leading-twist light-front wave functions (LFWFs), for the $q$-in-\textbf{P} valence-quark, are defined as follows\,\cite{Chouika:2017dhe,Chouika:2017rzs,Mezrag:2016hnp}:
\begin{subequations}
\label{eq:defLFWFs}
\begin{align}
    \psi_{\textbf{P}q}^{\uparrow \downarrow}(x,k_\perp^2;\zeta)=& \,\text{tr}\, Z_2\int_{k_\parallel}\delta_n^x\, \gamma_5 \gamma \cdot n \chi_{\textbf{P}}(k_-;P)\,,\label{eq:hel0def}\\    
    i k_{\perp\nu}\psi_{\textbf{P}q}^{\uparrow \uparrow}(x,k_\perp^2;\zeta)=& \,\text{tr}\,Z_2\int_{k_\parallel}\delta_n^x\, \gamma_5 n_\mu \sigma_{\mu\nu} \chi_{\textbf{P}}(k_-;P)\,.\label{eq:hel1def}
\end{align}
\end{subequations}
Here $\text{tr}$ indicates a trace over color and Dirac indices, $\delta_n^x:=\delta(n\cdot k - x\, n\cdot P)$ and $\int_{k_\parallel}=(1/\pi)\int dk_3 dk_4$. The Poincar\'e covariant Bethe-Salpeter wave function (BSWF) is denoted by $\chi_\textbf{P}$ and the kinematics is defined as usual: $x$ is the light-front fraction of the meson's total momentum carried by the $q$ valence-quark; $n$ is a light-like four-vector, such that $n^2=0$ and $n\cdot P = -m_\textbf{P}$. The labels $\uparrow \downarrow$ and $\uparrow \uparrow$ denote the helicity-0 and helicity-1 components, respectively. Finally, $\zeta$ is the resolving scale at which the LFWFs are defined and $Z_2$ the dressed-quark renormalization constant. 

Expressed as in Eqs.\,\eqref{eq:defLFWFs}, the LFWFs result from the projection of the Poincaré covariant BSWF onto the light-front. This ideas were first discussed in Ref.\,\cite{tHooft:1974pnl} and applied, for instance, in Ref.\,\cite{Chang:2013pq} to reconstruct the pion distribution amplitude (DA). As a result of such implementation, the LFWFs are expressed in a quasiparticle basis. Together with the restriction of the analysis to the leading-twist components, it is natural to adopt the hadronic scale $\zeta=\zeta_H$, at which all the properties of the hadron are fully described by its valence constituents. This approach has been proven robust in a variety of applications, e.g.\,\cite{Cui:2020tdf,Cui:2020dlm,Raya:2024glv,Raya:2022eqa,Raya:2021zrz,Zhang:2021mtn,Albino:2022gzs,Almeida-Zamora:2023bqb,Almeida-Zamora:2023rwg,Higuera-Angulo:2024oui,Albino:2026hhx,Xu:2018eii}. In this work, we aim to develop a robust and practical representation of the BSWF, leading to sensible profiles for the LFWFs. With the latter at hand, we gain access to an array of quantities that characterize the hadronic structure, including distribution amplitudes, distribution functions (DFs), electromagnetic and gravitational form factors. At a deeper level, these quantities are ultimately connected to GPDs and TMDs.

The formulation of the BSWF relies on the relation between the LFWF and the DAs and DFs, which are seemingly simpler objects depending on the single variable $x$. For instance, the leading-twist DA is given by:
\begin{equation}
    \label{eq:DAdef}
   f_\textbf{P} \varphi_\textbf{P}^q(x;\zeta_H)=\int \frac{d^2k_\perp}{16\pi^3} \psi_{\textbf{P}q}^{\uparrow \downarrow}(x,k_\perp^2;\zeta_H)\,,
\end{equation}
where $f_\textbf{P}$ is the meson's leptonic decay constant. On its part, the valence-quark DF connects as follows:
\begin{equation}
    \label{eq:DFdef}
   q_\textbf{P}(x;\zeta_H)=\int \frac{d^2k_\perp}{16\pi^3} \Big\{|\psi_{\textbf{P}q}^{\uparrow \downarrow}(x,k_\perp^2;\zeta_H)|^2+k_\perp^2 |\psi_{\textbf{P}q}^{\uparrow \uparrow}(x,k_\perp^2;\zeta_H)|^2\Big\}\,.
\end{equation}
The definitions in Eqs.\,\eqref{eq:DAdef} and\,\eqref{eq:DFdef} imply both DA and DF are unit-normalized. Moreover, owing to momentum conservation, their valence-antiquark counterparts can be readily obtained as:
\begin{equation}
    \varphi_\textbf{P}^{\bar{h}}(x;\zeta_H)=\,\varphi_\textbf{P}^q(1-x;\zeta_H)\,;\,
    \bar{h}_\textbf{P}(x;\zeta_H)=\,q_\textbf{P}(1-x;\zeta_H)\,.\label{eq:antiDF}
\end{equation}
Note for the DA, which evolves according to the ERBL equations,\,\cite{Efremov:1979qk,Lepage:1979zb,Lepage:1980fj}, this relation remains valid beyond $\zeta_H$. This is not the case for the DF, which instead follows DGLAP evolution\,\cite{Dokshitzer:1977sg,Gribov:1972ri,Lipatov:1974qm,Altarelli:1977zs}. Thus, guided by CSM determinations of these distributions,\,\cite{Cui:2020tdf,Cui:2020dlm}, and using the relations established in Eqs.\,\eqref{eq:defLFWFs}–\,\eqref{eq:DFdef}, we now proceed to construct the BSWF.

\section{Algebraic Models for LFWFs}
\label{sec:PTIR}

\subsection{Bethe-Salpeter Wave Function}
We begin by considering a pseudoscalar meson Bethe-Salpeter amplitude (BSA) expressed as follows:
\begin{equation}
\label{eq:BSAgen}
    \Gamma_{\textbf{P}}(k_-;P)=\gamma_5 \left[ i E_\textbf{P}(k_-;P)+\frac{\gamma \cdot P}{M_{q}+M_{\bar{h}}}F_\textbf{P}(k_-;P)\right]\;.
\end{equation}
Here $E_{\textbf{P}}$ and $F_{\textbf{P}}$ are scalar (dressing) functions depending on $k_-=k-P/2$, which denotes the relative momentum between the valence quark/antiquark, and the total bound-state's momentum $P^2=-m_\textbf{P}^2$. $M_{q,\bar{h}}$ labels the quark and antiquark constituent masses, defined in connection with the associated quark propagator:
\begin{equation}
\label{eq:prop}
    S_{q,\bar{h}}(p)= [-i \gamma \cdot p+M_{q,\bar{h}}]\,\Delta_{M_{q,\bar{h}}}(p^2)\,,\, \Delta_{M}^{-1}(s) := s+M^2\,.
\end{equation}

The meson BSA from Eq.\,\eqref{eq:BSAgen} retains two of the four amplitudes that characterize the complete object,\,\cite{Maris:1997tm}: the dominant amplitude $E_\text{P}$ and the pseudovector component $F_\text{P}$. Although subdominant, the latter ensures the correct asymptotic behavior of both the form factors,\,\cite{Maris:1998hc}, and, as we will see later, the LFWF. The omitted $P$-wave components are largely neglibible for the ground-state pseudoscalar, and their effects can be effectively absorbed into those amplitudes already considered\,\cite{Yao:2025xjx}.

Based upon Refs.\,\cite{Xu:2018eii,Zhang:2021mtn,Raya:2021zrz,Raya:2022eqa}, we employ a perturbation theory integral representation (PTIR) for the BSAs, such that ($\mathcal{A}_\textbf{P}=E_\textbf{P},F_\textbf{P}$):
\begin{equation}
\label{eq:BSAPTIR}
    n_\textbf{P}\mathcal{A}_\textbf{P}(k_-;P)=\int_{-1}^1 dw \,\rho_\textbf{P}^{\mathcal{A}}(w) \left[\Lambda_{\textbf{P}}^2 \,\Delta_{M_q}\left(k+\frac{w-1}{2}P\right)\right]\,,
\end{equation}
where $\rho_\textbf{P}$ denotes the associated spectral density, and $n_\textbf{P}$ the canonical normalization constant; $\Lambda_\textbf{P}$ is a mass scale to be determined. The corresponding BSWF follows straightforwardly:
\begin{equation}
\label{eq:BSWFgen}
    \chi_{\textbf{P}}(k_-;P)=S_{q}(k)\Gamma_{\textbf{P}}(k_-;P)S_{\bar{h}}(k-P)\,.
\end{equation}
From this point onward, unless otherwise specified, $\zeta_H$ will be assumed and the renormalization constants set to 1.

\subsection{PTIR: Leading BSA case}

For simplicity, let us first focus on the case $F_{\textbf{P}}(k_-;P)\equiv 0$. Considering the Mellin moments of the helicity-0 LFWF, Eq.\,\eqref{eq:hel0def}, a series of well-determined algebraic steps involving Feynman parametrization and change of variables yield:
\begin{align}
    \label{eq:Hel0E}
    \langle x^m\rangle_{\psi_{\textbf{P}q}^{\uparrow\downarrow}}&=\int_0^1 dx\,x^m\psi_{\textbf{P}q}^{\uparrow\downarrow}(x,k_\perp^2)\\
    &=12 \int_0^1 d\alpha \,\alpha^m [\alpha M_{\bar{h}}+(1-\alpha)M_q] \mathcal{X}_{\textbf{P}}^E(\alpha;\sigma_\perp^{2})\,,\notag
\end{align}
where we have defined $\sigma_\perp=k_\perp^2+\Omega_\textbf{P}^2$, and 
\begin{align}
\Omega_\textbf{P}^2 &= v M_q^2 + (1-v)\Lambda_\textbf{P}^2 
    + (M_{\bar{h}}^2 - M_q^2)\Big(\alpha - \tfrac{1}{2}(1-w)(1-v)\Big) \notag\\
    &\quad + \Big(\alpha(\alpha-1) + \tfrac{1}{4}(1-v)(1-w^2)\Big) m_\textbf{P}^2 \;,
    \label{eq:Omega}
\end{align}
in addition to the auxiliary function:
\begin{align}
    \label{eq:defXP}
    \mathcal{X}_{\textbf{P}}^{\mathcal{A}}(\alpha;\sigma_\perp^{1+\beta})&=\int_{w,v}\Bigg[\frac{\rho_\textbf{P}^\mathcal{A}(w)}{n_\textbf{P}}\frac{\Lambda_\textbf{P}^{2\beta}}{\sigma_\perp^{1+\beta}}\Bigg]\,,\\
    \int_{w,v}&:=\int_{-1}^{1-2\alpha}dw \int_{1+\frac{2\alpha}{w-1}}^1dv+\int_{1-2\alpha}^1dw \int_{\frac{w-1+2\alpha}{w+1}}^1dv\,. \notag
\end{align}
The unicity of the Mellin moments enable us to associate the Feynman parameter $\alpha$ with the physical variable $x$, and therefore, via Eq.\,\eqref{eq:Hel0E}, identify:
\begin{equation}
\label{eq:Psi0E}
    \psi_{\textbf{P}q}^{\uparrow\downarrow}(x,k_\perp^2)=\psi_{E}^{\uparrow\downarrow}(x,k_\perp^2):=12 \, \mathcal{X}_{\textbf{P}}^E(x;\sigma_\perp^{2}) \,[(1-x)M_q+x M_{\bar{h}}]\,.
\end{equation}
This result corresponds to the one derived in Refs.\,\cite{Zhang:2021mtn,Raya:2021zrz,Raya:2022eqa}. Proceeding analogously with the helicity-1 component, one arrives at ($j=1,2$):
\begin{equation}
\label{eq:Psi1E}
i k_{\perp j} \psi_{\textbf{P}q}^{\uparrow\uparrow}(x,k_\perp^2)=\psi_{E}^{\uparrow\uparrow}(x,k_\perp^2):=12\, \mathcal{X}_{\textbf{P}}^E(x;\sigma_\perp^{2})  k_{\perp j}\;.
\end{equation}
Clearly, when only the leading BSA is retained, both $\psi_{\textbf{P}q}^{\uparrow\downarrow}(x,k_\perp^2)$ and $\psi_{\textbf{P}q}^{\uparrow\uparrow}(x,k_\perp^2)$ display a $1/k_\perp^4$ power-law falloff. While such behavior is expected for the latter, the asympotic expectation for the helicity-0 piece is $1/k_\perp^2$,\,\cite{Lepage:1980fj}. In the present approach, as will be discussed below, this discrepancy is resolved by including the pseudo-vector component $F_{\textbf{P}}(k_-,P)$ in the meson BSA.

\subsection{PTIR: Complete BSA}
\label{sec:PTIRlfwf}

Accounting now for the dominant and the pseudovector components in the BSA, Eq.\,\eqref{eq:BSAgen}, then we find:
\begin{subequations}
\label{eq:Psi0F}
\begin{align}
\label{eq:Psi0Fa}
    \psi_{\textbf{P}q}^{\uparrow\downarrow}(x,k_\perp^2)=&\psi_{E}^{\uparrow\downarrow}(x,k_\perp^2)+\psi_{F}^{\uparrow\downarrow}(x,k_\perp^2)\,,\\
\label{eq:Psi0Fb}
    \psi_{F}^{\uparrow\downarrow}(x,k_\perp^2):=&\frac{6}{M_q+M_{\bar{h}}}\Bigg\{\mathcal{X}_{\textbf{P}}^F(x;\sigma_\perp)\Lambda_P^2\notag\\
    -&\mathcal{X}_{\textbf{P}}^F(x;\sigma_\perp^2)[2x(1-x)m_\textbf{P}^2+2M_q M_{\bar{h}}+\Omega_{\textbf{P}}^2]\Bigg\}\,.
\end{align}
\end{subequations}
Note that in the second term of $\psi_{F}^{\uparrow\downarrow}(x,k_\perp^2)$, $\Omega_\textbf{P}^2$ must be taken into account in the integration associated with $\mathcal{X}_{\textbf{P}}^F(x;\sigma_\perp^2)$, as it depends on its integration variables. Concerning the helicity-1 LFWF, the expressions are far more compact:
\begin{subequations}
\label{eq:Psi1F}
\begin{align}
\label{eq:Psi1Fa}
i k_{\perp j} \psi_{\textbf{P}q}^{\uparrow\uparrow}(x,k_\perp^2)=& i k_{\perp j} [\psi_{E}^{\uparrow\uparrow}(x,k_\perp^2)+\psi_{F}^{\uparrow\uparrow}(x,k_\perp^2)]\,\\
\label{eq:Psi1Fb}
i k_{\perp j} \psi_{F}^{\uparrow\uparrow}(x,k_\perp^2):
=& -12 \mathcal{X}_{\textbf{P}}^F(x;\sigma_\perp^{2})  k_{\perp j}\;.
\end{align}
\end{subequations}
Given Eqs.\,\eqref{eq:Psi0F} and\,\eqref{eq:Psi1F}, it is now clear that both $\psi_{\textbf{P}q}^{\uparrow\downarrow}$ and $\psi_{\textbf{P}q}^{\uparrow\uparrow}$ achieve both their proper asymptotic behaviors, i.e. $\sim1/k_\perp^{2}$ and $\sim1/k_\perp^4$, respectively. This feature is not properly captured in many comparable approaches\,\cite{Raya:2021zrz,Raya:2022eqa,Zhang:2021mtn,Mezrag:2016hnp,Albino:2022gzs,Almeida-Zamora:2023bqb,Almeida-Zamora:2023rwg,Higuera-Angulo:2024oui,Albino:2026hhx,Xu:2018eii,Chouika:2017dhe,Chouika:2017rzs}.


Motivated by the fact that the $E_\textbf{P}$ and $F_\textbf{P}$ BSAs exhibit an infrarred enhancement and monotonic decreasing with same asymptotic suppresion,\,\cite{Maris:1997tm}, we adopt the approximation:
\begin{equation}
    \rho_\textbf{P}^F(\omega):=\gamma_{\textbf{P}}\rho_{\textbf{P}}^E(w) \Rightarrow F_{\textbf{P}}(k_-;P)\propto E_{\textbf{P}}(k_-;P)\,,
\end{equation}
where $\gamma_\textbf{P}$ controls the relative strengh between the amplitudes. In such a case, the corresponding LFWFs can be expressed in a more compact manner:
\begin{align}
\notag
    \psi_{\textbf{P}q}^{\uparrow \downarrow}(x,k_\perp^2)=&12\Bigg\{
   \gamma_\textbf{P}\frac{\Lambda_\textbf{P}^2\,\mathcal{X}_{\textbf{P}}^E(x,\sigma_\perp)}{2(M_q+M_{\bar{h}})}\\ \notag
   +&\mathcal{X}_{\textbf{P}}^E(x,\sigma_\perp^2)\Bigg([(1-x)M_q+xM_{\bar{h}}] \\\label{eq:Psi0All}
    -& 
    \gamma_{\textbf{P}}\frac{[2x(1-x)m_{\textbf{P}}^2+2M_qM_{\bar{h}}+\Omega_\textbf{P}^2]}{2(M_q+M_{\bar{h}})}\Bigg) \Bigg\}\,,\\
\label{eq:Psi1All}
ik_{\perp j}\,\psi_{\textbf{P}q}^{\uparrow \uparrow}(x,k_\perp^2)=&12\,\mathcal{X}_{\textbf{P}}^E(x,\sigma_\perp^2)[1-\gamma_\textbf{P}]k_{\perp j}\,.
\end{align}
Notably, with the present considerations, if the relative strength between the considered BSAs is the same (namely $\gamma_\textbf{P}=1$), $\psi_{\textbf{P}}^{\uparrow \uparrow}(x,k_\perp)$ is identically zero. The omission of the two remaining BSAs causes no harm; as stated before, these are subdominant in nature and do not substantially modify the behavior of the LFWF. 

\begin{figure}[t]
\centerline{%
\begin{tabular}{c}
\includegraphics[width=0.4\textwidth]{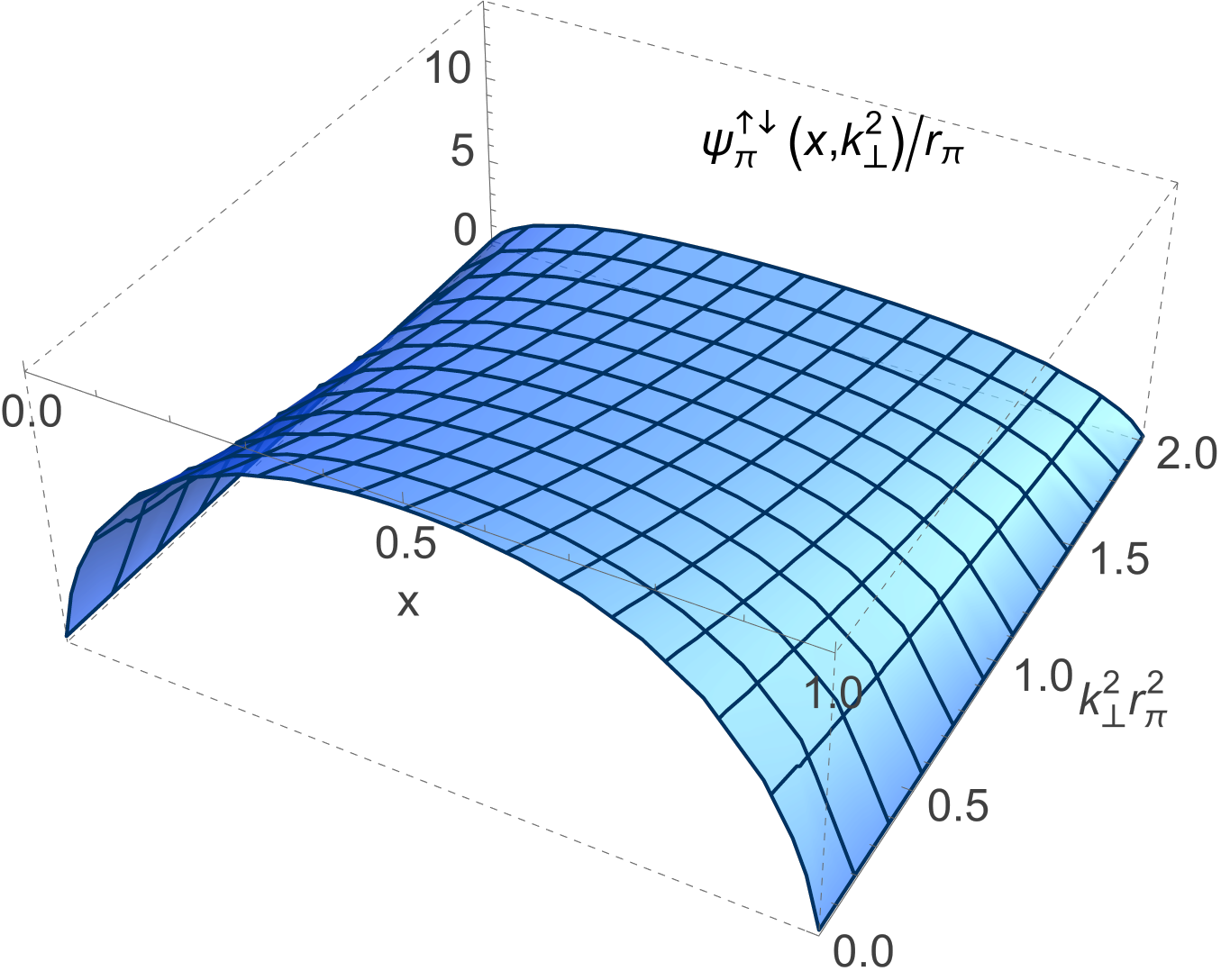} \\
\includegraphics[width=0.4\textwidth]{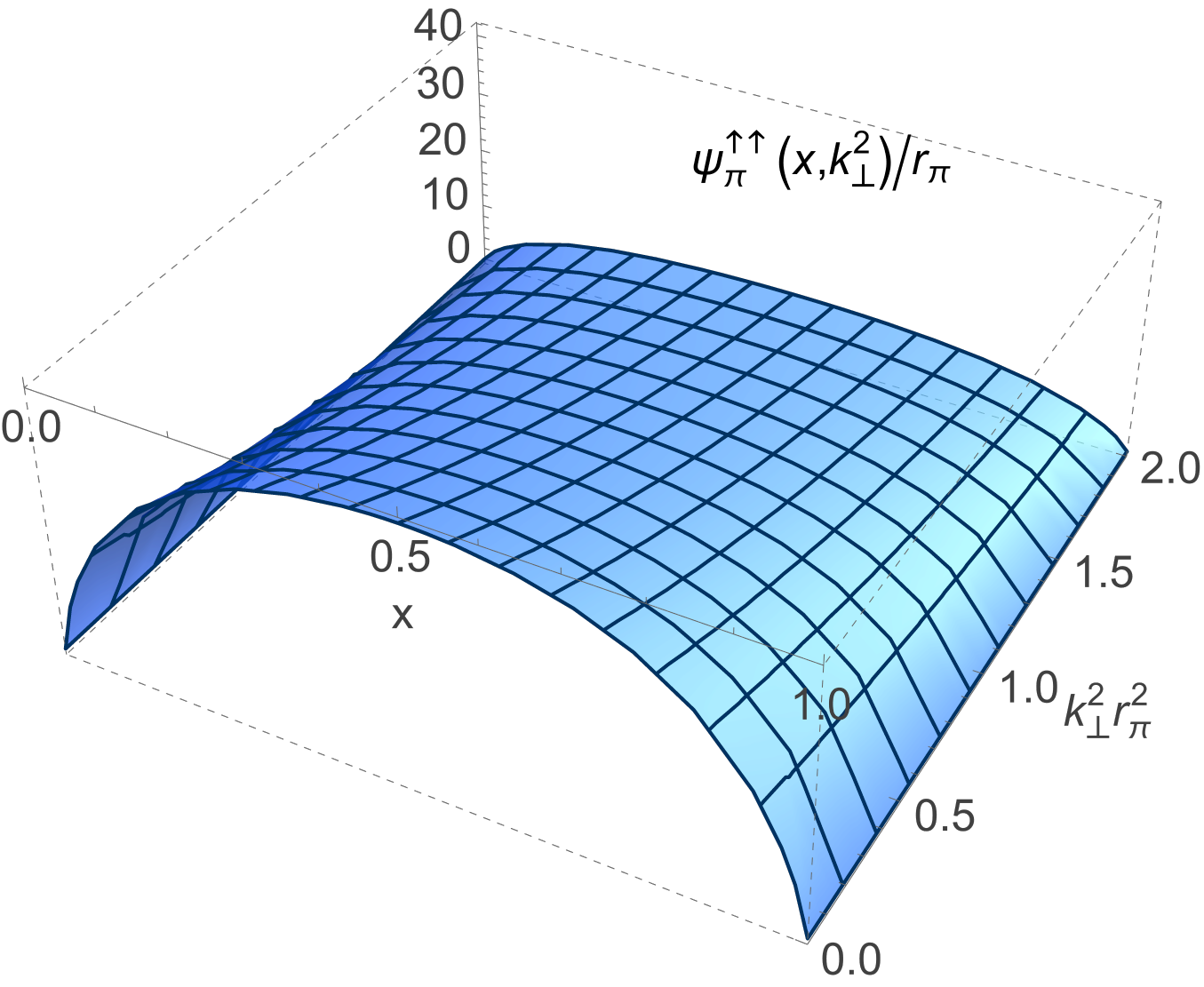}
\end{tabular}}
\caption{$\pi$ LFWFs as obtained form the PTIR approach described in Sec.\,\ref{sec:PTIRlfwf}, with the parameters from Table\,\ref{tab:PTIRparams}.}
\label{fig:LFWFs}     
\end{figure}

The PTIR model is thus complete once the mass scales ($m_\textbf{P}$, $M_{q,\bar{h}}$), the relative weight $\gamma_\textbf{P}$, and the spectral density $\rho_\textbf{P}(w)$ are determined. For the latter, we employ the flexible parametrization introduced in\,\cite{Xu:2018eii}:
\begin{align}
    \rho_\textbf{P}(w)=\frac{1+w\,v_\textbf{P}}{2a_\textbf{P}b_0^\textbf{P}}\Bigg[\text{sech}^2\Bigg(\frac{w-w_0^\textbf{P}}{2b_0^\textbf{P}}\Bigg)+\text{sech}^2\Bigg(\frac{w+w_0^\textbf{P}}{2b_0^\textbf{P}}\Bigg) \Bigg]\,.\label{eq:Spectral1}
\end{align}
The density's gross profile is controlled by $b_0^\textbf{P},\,w_0^\textbf{P}$, whereas $v_\textbf{P}\neq0$ modulates the asymmetry induced by the isospin-symmetry breaking;  the unit normalization is ensured via $a_\textbf{P}$. To fix the involved parameters, we consider the $\pi$ DA obtained in Refd.\,\cite{Cui:2020tdf,Cui:2020dlm}\,, parameterized according to\,\cite{Raya:2024ejx}:
\begin{equation}
\label{eq:DAparam}
    \varphi_\pi^q(x)=n_\varphi\ln\Bigg[1+\frac{x(1-x)}{\tilde{\rho}_{\pi}^2} \Bigg]\,,
\end{equation}
with $\tilde{\rho}_{\pi}=0.18$, and $n_\varphi$ ensuring the unit normalization. The best agreement is found with the parameters listed in Table\,\ref{tab:PTIRparams}, and the resulting LFWFs are shown in Fig.\,\ref{fig:LFWFs} .

\begin{table}[h!]
\centering
\caption{Used in Eqs.\,\eqref{eq:Psi0All}-\,\eqref{eq:Spectral1}, these parameters reproduce the pion DAs from Refs.\,\cite{Cui:2020dlm,Cui:2020tdf}, parameterized as in Eq.\,\eqref{eq:DAparam}. Here we consider the isospin symmetric limit $M_u=M_d$, and $\gamma_\textbf{P}\in[0,0.2]$. Mass units in GeV.}
\begin{tabular}{c | c  c c c c c c}
\hline
\textbf{P} & $m_\textbf{P}$ & $M_{u}$ & $M_{\bar{h}}$ & $\Lambda_\textbf{P}$ & $b_0^\textbf{P}$ & $w_0^\textbf{P}$ & $v_\textbf{P}$ \\
\hline
$\pi$ & $0.140$ & $0.325$ & $M_{u}$ & $M_{u}$ & $0.316$ & $1.15$ & $0$ \\
\hline
\end{tabular}
\label{tab:PTIRparams}
\end{table}

\subsection{Multipole Ansatz}
Consider the PTIR of the LFWFs in Eqs.\,\eqref{eq:Psi0All}–\eqref{eq:Psi1All}. The chiral limit is defined by $M_q=M_{\bar{h}}=\Lambda_{\textbf{P}}$ and $m_\textbf{P}=0$, such that $\Omega_\textbf{P}^2\to M_q^2$. As a result, the integration over $dv$ in Eq.\,\eqref{eq:defXP} becomes trivial, leaving only the integration over $dw$, which encodes the spectral density. Consequently, the $x$ and $k_\perp$ dependences decouple, indicating that any residual $x$–$k_\perp$ correlations originate from chiral and isospin symmetry breaking. In combination with Eq.\,\eqref{eq:DAdef}, the chiral limit permits to express the LFWFs as follows:
\begin{subequations}\label{eq:PsiCLDA}
\begin{align}
\label{eq:Psi0clDA}
    &\psi_{\textbf{P}q}^{\uparrow \downarrow}(x,k_\perp^2)\overset{c.l.}{=}
    \mathcal{N}_\varphi\Bigg[
    c_1 \frac{M_q^2}{(k_\perp^2+M_q^2)^2}
    +c_2 \frac{M_q^{2\delta}}{(k_\perp^2+M_q^2)^{1+\delta}}
    \Bigg]
    \varphi_\textbf{P}^q(x)\,,
    \\
\label{eq:Psi1clDA}
    &i k_{\perp j}\,\psi_{\textbf{P}q}^{\uparrow \uparrow}(x,k_\perp^2)\overset{c.l.}{=}
    \mathcal{N}_\varphi
    \left[
    c_3\frac{M_q^2}{(k_\perp^2+M_q^2)^2}
    \right]
    \frac{k_{\perp j}}{M_q}
    \varphi_\textbf{P}^q(x)\,,
\end{align}
\end{subequations}
Here $c_1=1-(3/4)\gamma_\textbf{P},\,c_2=\gamma_\textbf{P}/4,\,c_3=1-\gamma_\textbf{P}$; and $\delta$ is a positive definite parameter mimicking the effect of a renormalization-group logarithm and that enables the evaluation of the otherwise divergent integral associated with the DA, Eq.\,\eqref{eq:DAdef}. The normalization $\mathcal{N}_\varphi=16\pi^2f_\textbf{P}/(c_1+c_2/\delta)$ depends explicitly on $f_\textbf{P}$, as it immediately results from plugging Eq.\,\eqref{eq:Psi0clDA} into Eq.\,\eqref{eq:DAdef}.

On the other hand, when both helicity components are replaced in Eq.\,\eqref{eq:PDFdef} with Eqs.\,\eqref{eq:Psi0clDA} and \eqref{eq:Psi1clDA}, one is left with an explicit relation of the DF and the square of the DA, $q_\textbf{P}(x)\propto [\varphi_\textbf{P}^q(x)]^2$, established at the hadronic scale, which is well known in literature\,\cite{Cui:2020dlm,Cui:2020tdf}.


Then, in order of characterizing the $x-k_\perp$ correlations, that are expected to appear beyond the chiral limit, we introduce in Eq.\,\eqref{eq:PsiCLDA} the following replacement:

\begin{equation}
\label{eq:kappaDef}
    M_q\to \mathcal{M}_q:= \mathcal{M}_q(x)\,,
\end{equation}
where $\mathcal{M}_q(0)=M_q$. 
Eq.\,\eqref{eq:PDFdef} leads therefore to 

%
\begin{equation}\label{eq:qvsDA2kappa}
    q_\textbf{P}(x) = N_q 
    \frac{M_q^2}{\mathcal{M}_q^2(x)} [\varphi_\textbf{P}^q(x)]^2 \,,
\end{equation}
%
where $N_q$ is a dimensionless normalization guaranteeing baryon number conservation.

Thus, Eqs.\,\eqref{eq:PsiCLDA}, together with the replacement\,\eqref{eq:kappaDef}, 
define a multipole Ansatz (MA) for the pseudoscalar meson LFWFs that bypasses the need to explicitly specify a spectral density, instead making direct use of known information about the DF or DA. As discussed below, the MA also enables the derivation of a variety of algebraic relations and can be mapped onto the PTIR results for a suitable choice of the $x$-profile of $\mathcal{M}_q(x)$. As will become apparent, this profile is fixed once the DA and DF are specified. In the chiral limit, only one, either the DA or the DF, needs to be specified, as they become directly related by Eq.\,\eqref{eq:qvsDA2kappa} when $\mathcal{M}_q(x) \to M_q$. 

\section{Generalized Parton Distributions}
\label{sec:GPD}
The valence-quark generalized parton distribution (GPD) can be expressed in terms of the LFWFs via the overlap representation\,\cite{Diehl:2000xz,Burkardt:2002hr}:
\begin{align}
H^q_{\textbf{P}}(x,\xi,\Delta^2) =& \int \frac{d^2k_\perp}{16 \pi^3} \Bigg[
\psi^{\uparrow\downarrow\ast}_{\textbf{P}q}\left(x_-,\textbf{k}_{-}^2\right)  
\psi^{\uparrow\downarrow}_{\textbf{P}q}\left( x_+,\textbf{k}_{+}^2 \right) \notag \\
+&  (\textbf{k}_{-}\cdot \textbf{k}_{+})\, \psi^{\uparrow\uparrow\ast}_{\textbf{P}q}\left(x_-,\textbf{k}_{-}^2\right)  
\psi^{\uparrow\uparrow}_{\textbf{P}q}\left( x_+,\textbf{k}_{+}^2 \right)\Bigg]\,.
\label{eq:overlap}
\end{align} 
Here $x_\pm=(1\pm\xi)/(1\pm\xi)$, $\textbf{k}_{\mp}=k_{\perp}\pm(\Delta_\perp/2)(1-x)/(1\mp\xi)$ and $\Delta_\perp^2=\Delta^2(1-\xi^2)-4\xi^2m_\textbf{P}^2$. The quantity $\Delta$ denotes the momentum transfered by the probe and $\xi$ is the skewness variable. The overlap representation ensures the so called positivity,\,\cite{Pire:1998nw}, and holds in the DGLAP region, corresponding to $|x|\geq\xi$. Furthermore, owing to time-reversal invariance, $H_\textbf{P}^q(x,\xi,\Delta^2)=H_\textbf{P}^q(x,-\xi,\Delta^2)$, so we only consider $\xi\geq0$ in the following. The associated DF is recovered in the forward limit:
\begin{equation}
    \label{eq:PDFdef}
    q_\textbf{P}(x)=H_\textbf{P}^q(x,0,0)\,,
\end{equation}
which corresponds to the result anticipated by Eq.\,\eqref{eq:DFdef}. 

A given valence-quark contrbution to the meson elastic electromagnetic form factor stems from the zeroth moment of the GPD, namely:
\begin{equation}
    \label{eq:EFFdef}
    F_\textbf{P}^q=\int_{-1}^1 dx\,H_\textbf{P}^q(x,\xi,\Delta^2)\,.
\end{equation}
This quantity is independent of $\xi$ and one can simply take $\xi=0$. The complete meson EFF is obtained by adding up the individual quark contributions:
\begin{equation}
    \label{eq:EFFtotal}
    F_\textbf{P}(\Delta^2)=e_q F^q_\textbf{P}(\Delta^2)+e_{\bar{h}}F^{\bar{h}}_\textbf{P}(\Delta^2)\,,
\end{equation}
where $e_q,e_{\bar{h}}$ are the valence-constituent electric charges in units of the positron charge. The charge radius is defined as usual:
\begin{align}
(r_{E}^{\textbf{P}})^2=\,e_q (r_{E}^{\textbf{P}\bar{h}})^2+e_{\bar{h}}(r_{E}^{\textbf{P}\bar{h}})^2\,;\,\label{eq:radii}
    (r_{E}^{\textbf{P}q,\bar{h}})^2=-6\frac{\partial F_\textbf{P}^{q,\bar{h}}(\Delta^2)}{\partial\Delta^2}\Bigg|_{\Delta^2=0}\,.
\end{align}
If the valence-quark and antiquark are assummed to be mass degenerate, then we take:
\begin{equation}
    F_\textbf{P}(\Delta^2)=F_\textbf{P}^q(\Delta^2)\,,\,r_{E}^\textbf{P}=r_E^{\textbf{P}q}\,.
\end{equation}
The gravitational form factors can be identified from the first moment of the GPD:
\begin{equation}
\label{eq:GFFdef}
    \theta_2^{\textbf{P}q}(\Delta^2)+\xi^2 \theta_1^{\textbf{P}q}(\Delta^2) = \int_{-1}^1dx \,xH_{\textbf{P}}^q(x,\xi,\Delta^2)\,.
\end{equation}
Working at $\zeta_H$, the meson’s total GFFs are obtained from each valence-quark contribution:
\begin{equation}
    \theta_{1,2}^\textbf{P}(\Delta^2)=\theta_{1,2}^{\textbf{P}q}(\Delta^2)+\theta_{1,2}^{\textbf{P}\bar{h}}(\Delta^2)\,.
\end{equation}
The form factor $\theta_2^\textbf{P}(\Delta^2)$ is associated with the mass distribution inside the meson, whereas $\theta_1^\textbf{P}(\Delta^2)$ encodes the internal pressures. Adapting the usual expression for form factor radii\,\footnote{Note that at the hadronic scale, $\theta_2^{\textbf{P}q}(0)$ is merely the first Mellin moment of the DF, namely $\langle x\rangle_q=\int_0^1dx\,q_\textbf{P}(x)$.}:
\begin{equation}
    \label{eq:MassRadDef}
    [r_{\theta_2}^{\textbf{P}}]^2=
    \theta_2^{\textbf{P}q}(0)[r_{\theta_2}^{\textbf{P}q}]^2+\theta_2^{\textbf{P}\bar{h}}(0)
    [r_{\theta_2}^{\textbf{P}\bar{h}}]^2\,.
\end{equation}
So, in the isospin-symmetric limit,
\begin{equation}
    \label{eq:radiitheta12iso}
    \theta_{1,2}^\textbf{P}(\Delta^2)=2\theta_{1,2}^{\textbf{P}q}(\Delta^2)\,,\,r^\textbf{P}_{{\theta_{1,2}}}=r^{\textbf{P}q}_{{\theta_{1,2}}}\,.
\end{equation}
It is worth noting that $\theta_1(\Delta^2)$ lies, in principle, beyond the scope of the overlap representation, as it is related to the so-called $D$-term ambiguity \cite{Mezrag:2022pqk,Polyakov:2018zvc}. Hence, despite existing strategies to extend the DGLAP GPD into the ERBL region (e.g. Refs.\,\cite{Chouika:2017dhe,Chouika:2017rzs,Chavez:2021llq,DallOlio:2024vjv}), the discussion on $\theta_1(\Delta^2)$ is omitted in the present work. Instead, we concentrate on quantities that are accessible within the DGLAP kinematic domain of the GPD.

\subsection{GPDs: PTIR model}
\begin{figure}[t]
\centerline{
\begin{tabular}{c}
\includegraphics[width=0.45\textwidth]{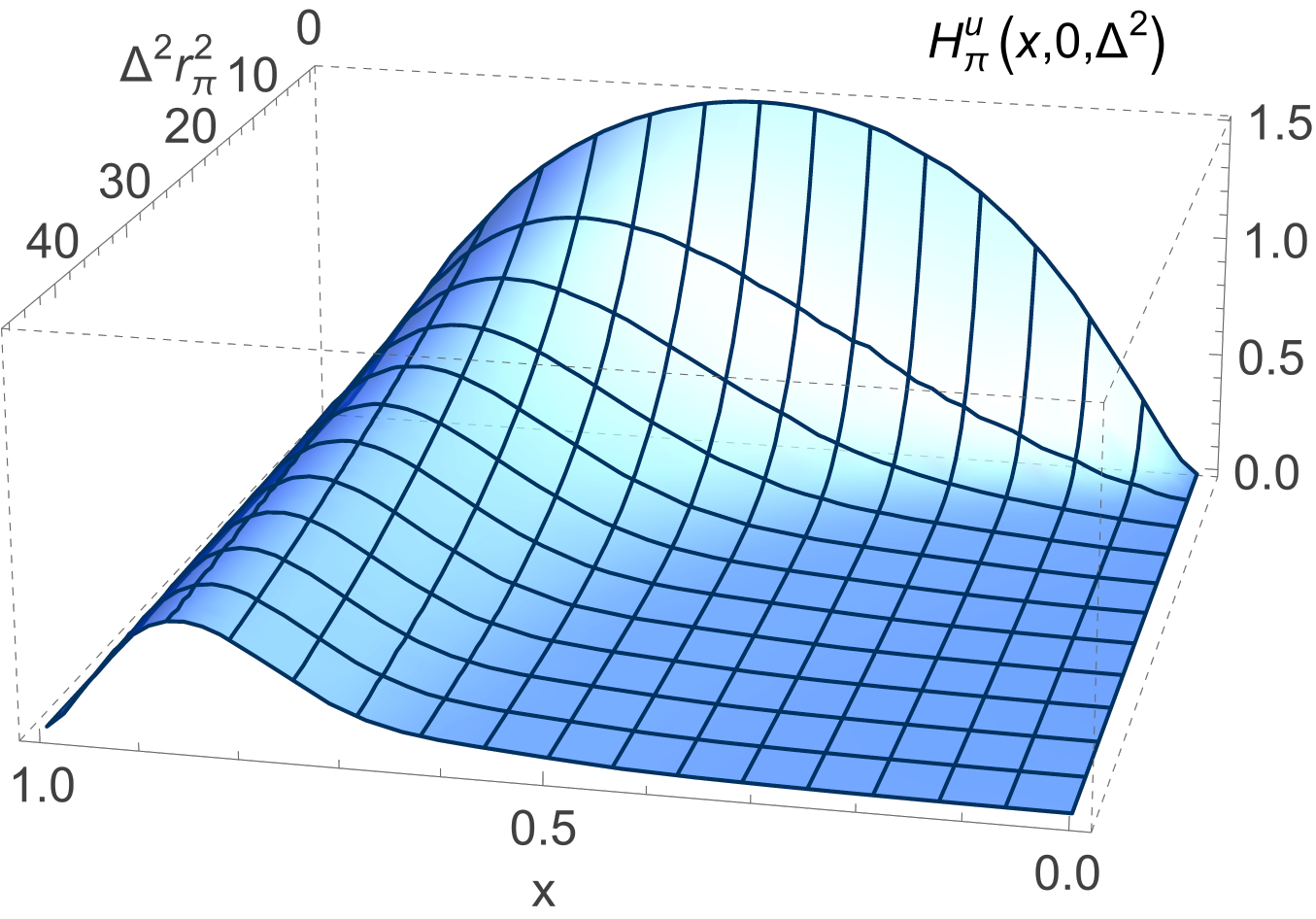} 
\end{tabular}
}
\caption{$\pi$ GPD obtained via the overlap representation, employing the PTIR.}
\label{fig:GPDsPTIR} 
\end{figure}

Employing the parameter set of Table\,\ref{tab:PTIRparams}, the pion valence-quark GPD derived through the overlap representationis displayed in Fig\,\ref{fig:GPDsPTIR}. One observes that $H_\pi^u(x,0,0)$ is broad and symmetrical around $x=1/2$ and, as the momentum transfer increases, the maxima shift towards $x\to1$, indicating the domain in which the struck valence quark carries nearly all of the hadron’s longitudinal momentum.

\begin{figure}[t]
\centerline{
\begin{tabular}{c}
\includegraphics[width=0.45\textwidth]{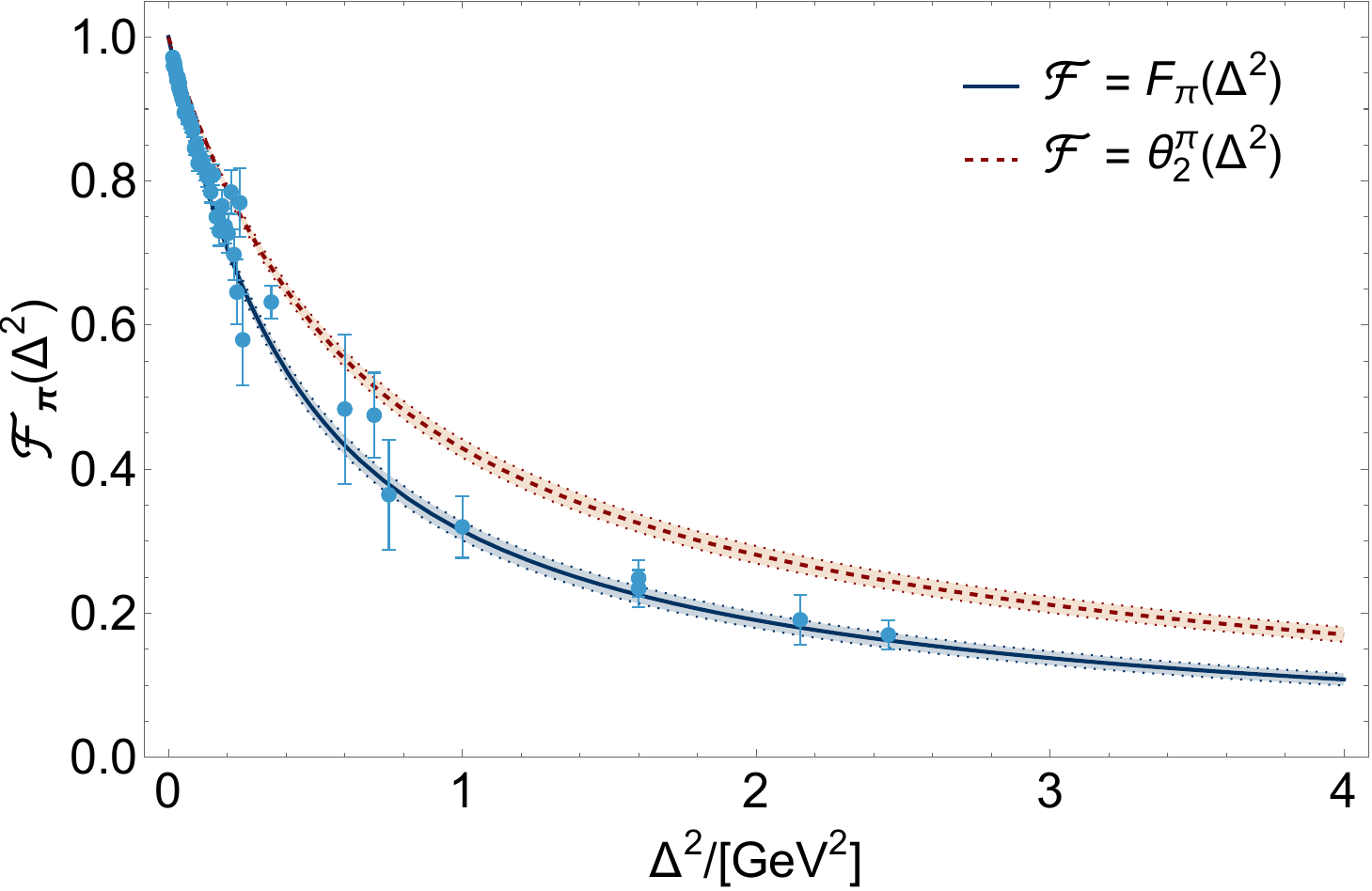}
\end{tabular}
}
\caption{Pion electromagnetic and $\theta_2^\pi(\Delta^2)$ gravitational form factors. The error reflects the variation on $\gamma_\pi\in[0,0.2]$, which weights the pseudovector component in the meson BSA. Experimental data points associated to $F_\pi(\Delta^2)$ are taken from Refs.\,\cite{NA7:1986vav,Horn:2007ug,JeffersonLab:2008jve}.}
\label{fig:piFFs} 
\end{figure}

The corresponding pion electromagnetic and $\theta_2^\pi(\Delta^2)$ gravitational form factors are displayed in Fig\,\ref{fig:piFFs} . Notably, $F_\pi(\Delta^2)$ shows excellent agreement with the available experimental data. On its part, $\theta_2^\pi(\Delta^2)$ manifests a less pronounced $\Delta^2$ falloff. This pattern is consistent with both theoretical and phenomenological expectations\,\cite{Xu:2023bwv,Xu:2023izo,Yao:2024ixu,Yao:2025fnb,Wang:2024fjt,Hackett:2023nkr,Kumano:2017lhr}, and results in a more compressed mass distribution than that associated with the charge\,\cite{Raya:2024ejx,Raya:2022eqa,Raya:2024glv}. A measure of this compression is provided via the corresponding charge and mass radii:
\begin{equation}
\label{eq:RadiiValsPi}
    r_E^{\pi}=0.66(1)\,\text{fm}\,,
    r_{\theta_2}^{\pi}=0.56(1)\,\text{fm}\,.
\end{equation}
These values are in agreement with empirical extractions\,\cite{ParticleDataGroup:2024cfk,Xu:2023bwv,Cui:2022fyr}, and imply $(r_{\theta_2}^\pi/r_E^\pi)^2=0.58(4)$, consistent with the so-called physical bounds\,\cite{Xu:2023bwv}. 

\subsection{GPDs: Multipole Ansatz}
Employing the LFWFs from Eqs.\,\eqref{eq:PsiCLDA}, supplemented with the prescription\,\eqref{eq:kappaDef} and Eq.\,\eqref{eq:qvsDA2kappa} to replace the DA with the DF, 
the overlap representation in Eq.\,\eqref{eq:overlap} yields:
\begin{equation}
    \label{eq:GPDtot0}
    H_\textbf{P}^{q}(x,\xi,\Delta^2)=\sqrt{q_\textbf{P}(x^+)q_\textbf{P}(x^-)}\, \Phi_{\textbf{P}q}(\hat{z})\,,
\end{equation}
with the function $\Phi_{\textbf{P}q}(\hat{z})$ controling the off-forward behavior of the distribution, such that:
\begin{subequations}
    \label{eq:GPDtot}
\begin{align}
    \Phi_{\textbf{P}q}(\hat{z})=&\,\mathcal{N}_H\left[\Phi_{\textbf{P}q}^{\uparrow\downarrow}(\hat{z})+\Phi_{\textbf{P}q}^{\uparrow\uparrow}(\hat{z})\right]\label{eq:GPDphi}\,,\\
    \Phi_{\textbf{P}q}^{\uparrow\downarrow}(\hat{z})=&\Bigg[2c_1^2 I_{1,1,3}+c_2^2 \frac{\Gamma(1+2\delta)}{\Gamma^2(1+\delta)}I_{\delta,\delta,1+2\delta}(\hat{z})\label{eq:Hel0GPD}\\
    +&c_1c_2(1+\delta)[I_{\delta,1,2+\delta}(\hat{z})+I_{1,\delta,2+\delta}(\hat{z})]
    \Bigg]\,,
    \notag\\
    \Phi_{\textbf{P}q}^{\uparrow\uparrow}(\hat{z})=& \left[I_{1,1,2}(\hat{z}) - 2\hat{z}\, I_{2,2,3}(\hat{z})\right]c_3^2
    \,;
    \label{eq:Hel1GPD}
\end{align}
\end{subequations}
and $\mathcal{N}_H$ guaranteeing baryon number conservation in the forward limit, \emph{i.e.}, $\Phi_{\textbf{P}q}(0)=1$, and then
\begin{equation}
\label{eq:normLFWF}
\mathcal{N}_H:= \frac{6}{2c_1^2+c_3^2+\frac{12 c_1 c_2}{2+\delta}+\frac{6c_2^2}{1+2\delta}}\,.
\end{equation}
In addition, we have adopted the definitions:
\begin{subequations}
\label{eq:IntsI}
\begin{align}
\label{eq:IntsIa}
    I_{i,j,k}(\hat{z}):=\int_0^1du \frac{u^i(1-u)^j}{(1+u(1-u)\hat{z})^k}\,,
    \\
    \mathrm{with}\;\;\;\;\hat{z}:=\frac{\Delta_\perp^2}{\mathcal{M}_q^2(x)}\frac{(1-x)^2}{(1-\xi^2)^2}\;.
\label{eq:IntsIb}
\end{align}
\end{subequations}
This sort of integrals can be evaluated for the required cases, and the function $\mathcal{M}(x)$, which is an expression of the $x-k_\perp$ correlations in the LFWFs, 
reshapes the parameter $\hat{z}$ in Eq.\,\eqref{eq:IntsIa} as shown by \eqref{eq:IntsIb}, and hence the argument of the function $\Phi_{\mathbf{P}q}$ introduced in Eq.\,\eqref{eq:GPDtot0} for the GPD. 

Having expressed the GPD as in Eq.\,\eqref{eq:GPDtot0}, it is clear that the off-forward part of the GPD depends only on the single kinematic variable $\hat{z}$. A series of corollaries follow from such representation\,\cite{Raya:2021zrz,Raya:2022eqa,Zhang:2021mtn,Xu:2023bwv}. For instance, set $\xi=0$ and adopt the limit\footnote{
A non-zero, positive-definite $\delta$ is needed to avoid a non-physical singularity in the definition of the DA, Eq.\,\eqref{eq:DAdef}; it can be, however, very small and set to zero as an approximation, only when considering the evaluation of the GPD and the PDF in the forward limit. A different assumption, often made in the literature, \emph{e.g.}, in Refs.\,\cite{Mezrag:2016hnp,Chouika:2017rzs,Zhang:2021mtn,Raya:2021zrz}, is to take $c_2\to 0$, effectively implying the neglect of the LFWF contribution from the BSA component $F_\mathbf{P}$. 
}
$\delta\to0$. The expansion around $\hat{z}\approx 0$ yields:
\begin{equation}
\label{eq:Phinear0}
    \Phi_{\textbf{P}q}(\hat{z}\approx0)=1-\frac{1}{10}\frac{160-232\gamma_\textbf{P}+93\gamma_\textbf{P}^2}{24+(11\gamma_\textbf{P}-28)\gamma_\textbf{P}}\hat{z}+\mathcal{O}(\hat{z}^2)\,.
\end{equation}
Then, given Eq.\,\eqref{eq:radii}, one can readily identify:
\begin{align}
   \label{eq:PhiandChargeMass}
    \frac{1}{10\,M_q^2}\frac{160-232\gamma_\textbf{P}+93\gamma_\textbf{P}^2}{24+(11\gamma_\textbf{P}-28)\gamma_\textbf{P}}=\frac{(r_E^{\textbf{P}q})^2}{6\langle (1-x)^2 \rangle_{\hat{q}}}\,=\frac{\theta_2^{\textbf{P}q}(0)(r_{\theta_2}^{\textbf{P}q})^2}{6\langle x(1-x)^2 \rangle_{\hat{q}}}\,,     
\end{align}
where the Mellin moments $\langle x^n \rangle_{\hat{q}}$ are defined as:
\begin{equation}
    \label{eq:defqhat}
    \langle x^n \rangle_{\hat{q}}=\int_0^1dx\,x^n\,q_\textbf{P}(x)\frac{M_q^2}{\mathcal{M}_q^2(x)}\,.
\end{equation}
The mass-to-charge ratio for the valence-quark $q$ is thus expressed in terms of hadronic scale moments:
\begin{equation}
    \theta_2^{\textbf{P}q}(0)\left(\frac{r_{\theta_2}^{\textbf{P}q}}{r_E^{\textbf{P}q}}\right)^2=\frac{\langle x(1-x)^2 \rangle_{\hat{q}}}{\langle (1-x)^2 \rangle_{\hat{q}}}\,,
\end{equation}
while for meson's total radii:
\begin{equation}
    \label{eq:MasstoCharge}
    \left(\frac{r_{\theta_2}^\textbf{P}}{r_E^\textbf{P}}\right)^2=\frac{\theta_2^{\textbf{P}q}(0)\langle x(1-x)^2\rangle_{\hat{q}} + \theta_2^{\textbf{P}\bar{h}}(0)\langle x^2(1-x)\rangle_{\hat{q}}}{e_q\langle (1-x)^2\rangle_{\hat{q}} + e_{\bar{h}}\langle x^2\rangle_{\hat{q}}}\,.
\end{equation}
In the isospin-symmetric limit, we meet the expectatives from Refs.\,\cite{Raya:2021zrz,Raya:2022eqa,Zhang:2021mtn}, namely:
\begin{equation}
    \label{eq:MasstoChargeSym}
    \left(\frac{r_{\theta_2}^\textbf{P}}{r_E^\textbf{P}}\right)^2=\frac{2\langle x^2(1-x)\rangle_q}{\langle x^2\rangle_q}\,.
\end{equation}
Turning now to the chiral limit, two cases stand out:
\begin{subequations}
    \label{eq:radiigamma0}
\begin{align}
\label{eq:Radiusgamma0hel0}
    \frac{3}{5M_q^2}=&\frac{r_\textbf{P}^2}{6\langle x^2 \rangle_q}\,\,\,\,(\text{for}\,\gamma_\textbf{P}\equiv0\,,\,\psi_{\textbf{P}}^{\uparrow\uparrow}\equiv0)\,,\\
    \label{eq:Radiusgamma0hel1}
    \frac{2}{3M_q^2}=&\frac{r_\textbf{P}^2}{6\langle x^2 \rangle_q}\,\,\,\,(\text{for}\,\gamma_\textbf{P}\equiv0\,)\,.
\end{align}
\end{subequations}
The result in Eq.\,\eqref{eq:Radiusgamma0hel0} recovers the chiral limit expectations from Refs.\,\cite{Albino:2022gzs,Raya:2021zrz,Zhang:2021mtn,Raya:2022eqa}, while Eq.\,\eqref{eq:Radiusgamma0hel1} corresponds to a generalization of the findings in Refs.\,\cite{Chouika:2017dhe,Chouika:2017rzs} to arbitrary DA/DF. This implies that, for the same constituent mass and DF, inclusion of the helicity-1 contributions leads to a modest increase in the charge radius, and a slightly more pronounced $\Delta^2$ falloff. 

\begin{figure}[t]
\centerline{
\begin{tabular}{c}
\includegraphics[width=0.45\textwidth]{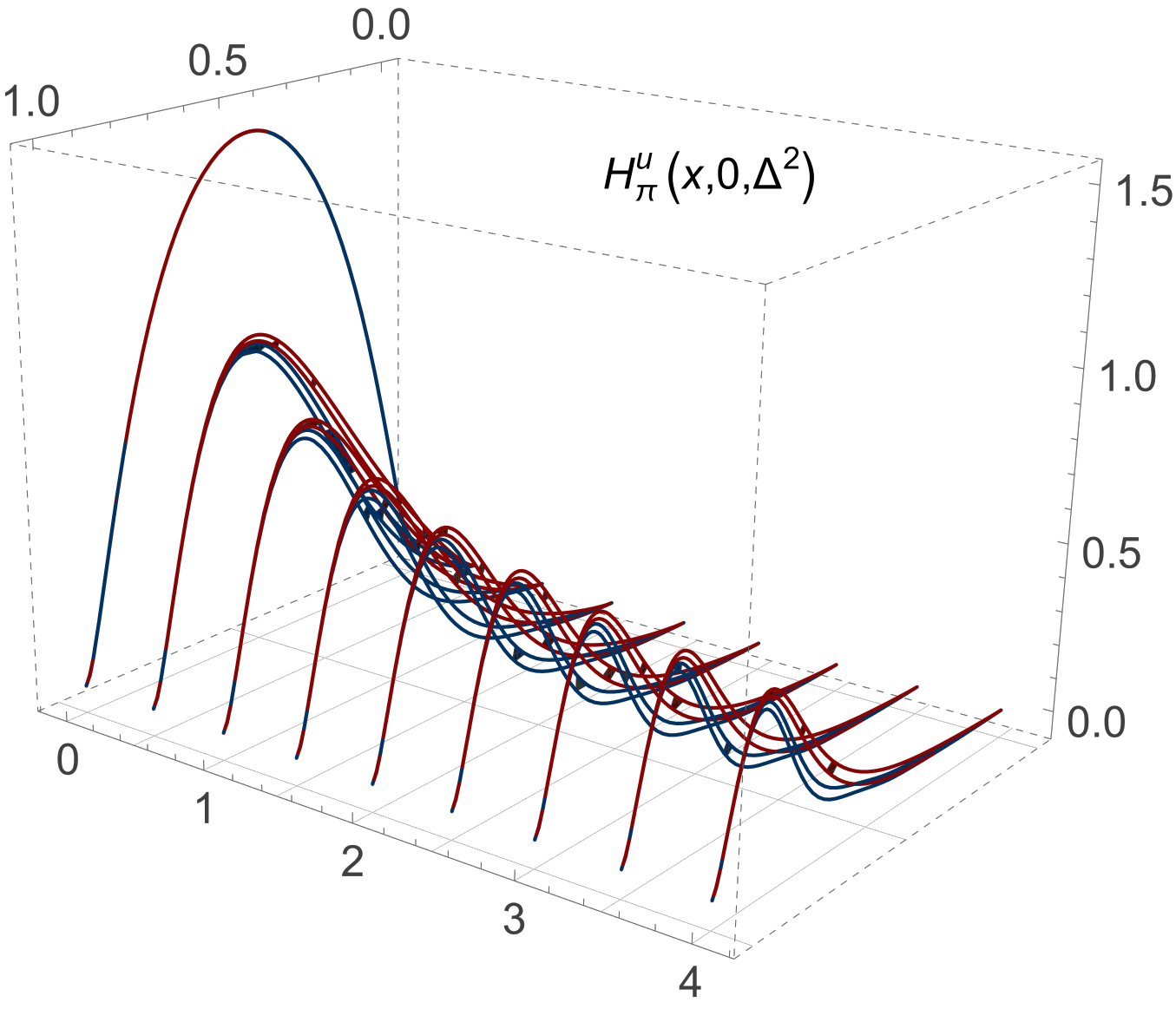}
\end{tabular}
}
\caption{$\pi$ GPDs obtained via PTIR (Blue) and through the MA (Red).}
\label{fig:GPDmodels} 
\end{figure}

\begin{figure}[t]
\centerline{
\begin{tabular}{c}
\includegraphics[width=0.45\textwidth]{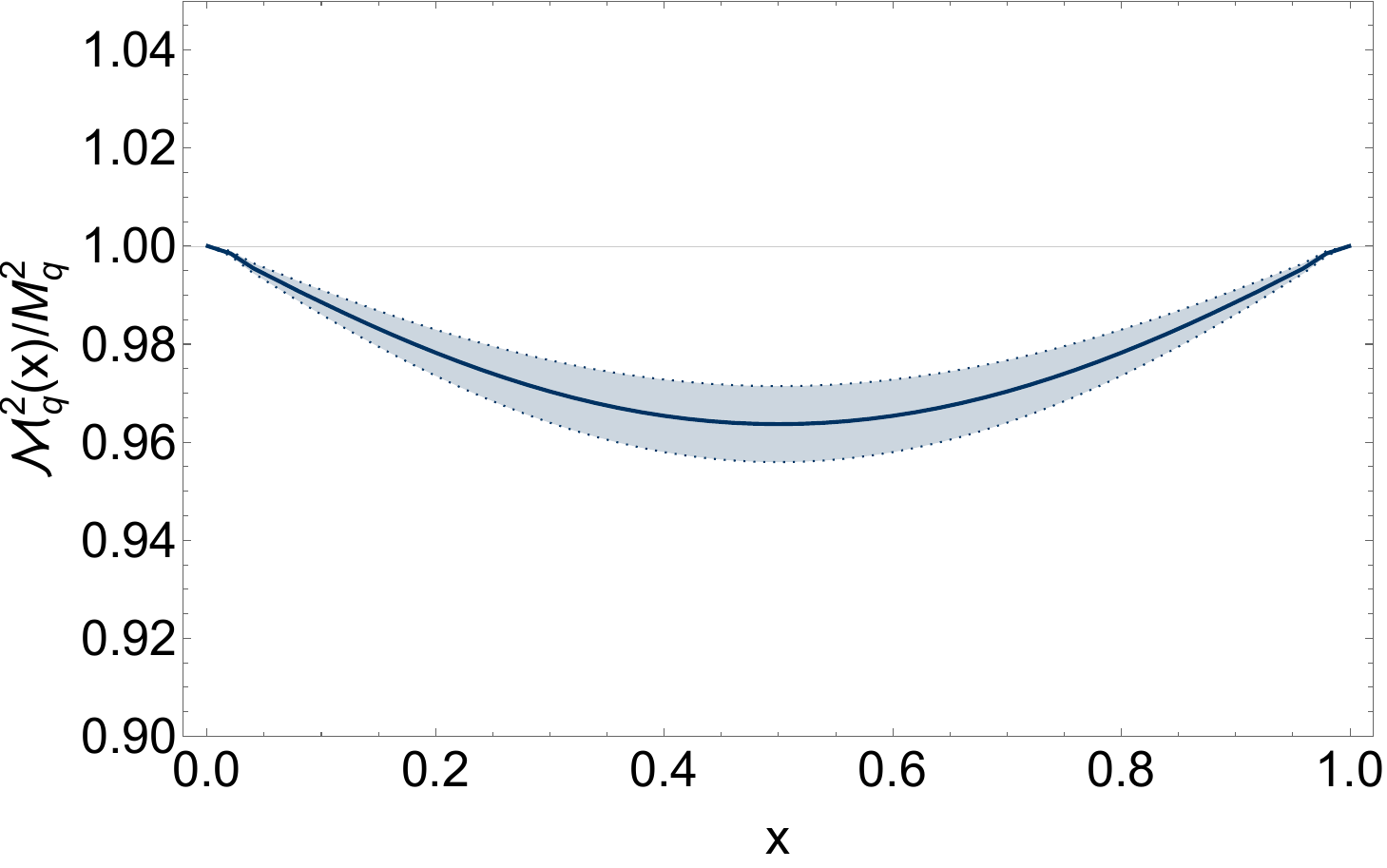}
\end{tabular}
}
\caption{$\pi$ profile function $\mathcal{M}^2_q(x)$, introduced in Eq.\,\eqref{eq:kappaDef}. The upper (lower) bound corresponds to $\gamma_\pi= 0.0\,(0.2)$.}
\label{fig:Kappa} 
\end{figure}

In Fig.\,\ref{fig:GPDmodels}, we compare the PTIR and MA results on the pion GPD. The same inputs are used in each case. The profile adopted by $\mathcal{M}_q^2(x)$ is shown in Fig\,\ref{fig:Kappa}. It is evident that the PTIR and MA results are in mutual agreement and both lie close to the factorized limit. This behavior can be attributed to the fact that deviations from $x-k_\perp$ factorization are controlled by the degree of explicit chiral symmetry breaking, making the factorized approximation particularly reliable for the pion.

\section{Transverse Structure and Dynamics}
\label{sec:Transverse}
Let us now capitalize on the pion  transverse structure as revealed by the meson GPD in the impact parameter space (IPS-GPD), the associated charge and mass density profiles, and the corresponding TMDs.

\subsection{GPDs in Impact Parameter Space}
The impact parameter space GPD (IPS-GPD) is obtained by evaluating a two-dimensional Fourier transform\,\cite{Diehl:2002he,Burkardt:2000za}:
\begin{equation}
\label{eq:IPSdef}
    q_\textbf{P}(x,|b_\perp|)=\int_0^\infty \frac{d\Delta}{2\pi}\Delta J_0(|b_\perp|\Delta)\,H_\textbf{P}^q(x,0,\Delta^2)\,,
\end{equation}
where $J_0$ is a cylindrical Bessel function. This density encodes the probability of finding the valence-quark $q$, carrying a longitudinal momentum fraction $x$, at a transverse distance $|b_\perp|$ from the meson's centre of transverse momentum. Being defined at $\xi=0$, the IPS-GPD is completely determined by the properties of the GPD on the DGLAP domain.

Acknowledging the strong agreement between the PTIR and MA approaches, and to exploit the algebraic advantages of the latter, we consider the GPD representation in Eq.\,\eqref{eq:GPDtot}. The IPS-GPD is then conveniently expressed as follows:
\begin{equation}
\label{eq:IPSMA}
    q_\textbf{P}(x,|b_\perp|)=\frac{q_\textbf{P}(x)\mathcal{M}_q^2(x)}{(1-x)^2}\int_0^\infty \frac{\hat{s}d\hat{s}}{2\pi}\Phi_{\textbf{P}q}(\hat{s}^2)J_0\left(\frac{|b_\perp| \hat{s} }{1-x} \sqrt{\mathcal{M}_q^2(x)}\right)\,.
\end{equation}
For the factorized case, $\mathcal{M}_q^2(x)\to M_q^2$, the maximum of the distribution lies at $q_\textbf{P}(x\to1,|b_\perp|\to0)$. Since Eq.\,\eqref{eq:IPSMA} defines a rotationally invariant quantity, it is often convenient to consider:
\begin{equation}
\label{eq:defIP}
    \mathcal{I}_\textbf{P}^q(x,|b_\perp|):=2\pi |b_\perp|\,q_\textbf{P}(x,|b_\perp|)\,.
\end{equation}
The peak of this distribution is shifted to $|b_\perp|>0$, by an amount that reflects features of the internal dynamics\,\cite{Raya:2021zrz,Raya:2022eqa,Zhang:2021mtn}. Beyond the factorized case, $\mathcal{M}_q^2(x)$ introduces additional modulations. 

One could also consider the longitudinal light-front distribution of the mean-squared transverse extent (MSTE):
\begin{equation}
\label{eq:MSTE}
    \langle b_{\perp}^2(x)\rangle_\textbf{P}^q:=\int_0^\infty |b_\perp|^2 \mathcal{I}_\textbf{P}^q(x,|b_\perp|) = -4\frac{\partial}{\partial \Delta^2}H_\textbf{P}^q(x,0,\Delta^2)\Big|_{\Delta^2=0}\;.
\end{equation}
In the present case, a simple expression for this quantity is obtained:
\begin{align}
\label{eq:MSTEMA}
    \langle b_\perp^2(x)\rangle_{\textbf{P}}^q=-4\,q_\textbf{P}(x)\frac{\partial \Phi_\textbf{P}(\hat{z})}{\partial\hat{z}}\Big|_{\hat{z}=0}=\frac{2}{3}(r_E^{\textbf{P}q})^2\frac{(1-x)^2q_\textbf{P}(x)}{\langle(1-x)^2\rangle_{\hat{q}}}\frac{M_q^2}{\mathcal{M}_q^2(x)}\,,
\end{align}
where the last equal sign follows from Eq.\,\eqref{eq:PhiandChargeMass}.

The associated expectation value follows straightforwardly:
\begin{equation}
\label{eq:MSTEave}
    \langle |b_{\perp}|^2\rangle_\textbf{P}^q:=\int_0^1dx\, \langle b_{\perp}^2(x)\rangle_\textbf{P}^q\,=\frac{2}{3}(r_E^{\textbf{P}q})^2\,.
\end{equation}
This result is more general and follows from the dimensionality in which the charge radius and $\langle |b_{\perp}|^2\rangle_\textbf{P}$ are defined. Moreover,
this quantity amounts to two-thirds of its corresponding contribution to the meson’s charge radius squared, with the latter coinciding with the total charge radius of the hadron in the isospin-symmetric limit.

\begin{figure}[t]
\centerline{
\begin{tabular}{c}
\includegraphics[width=0.45\textwidth]{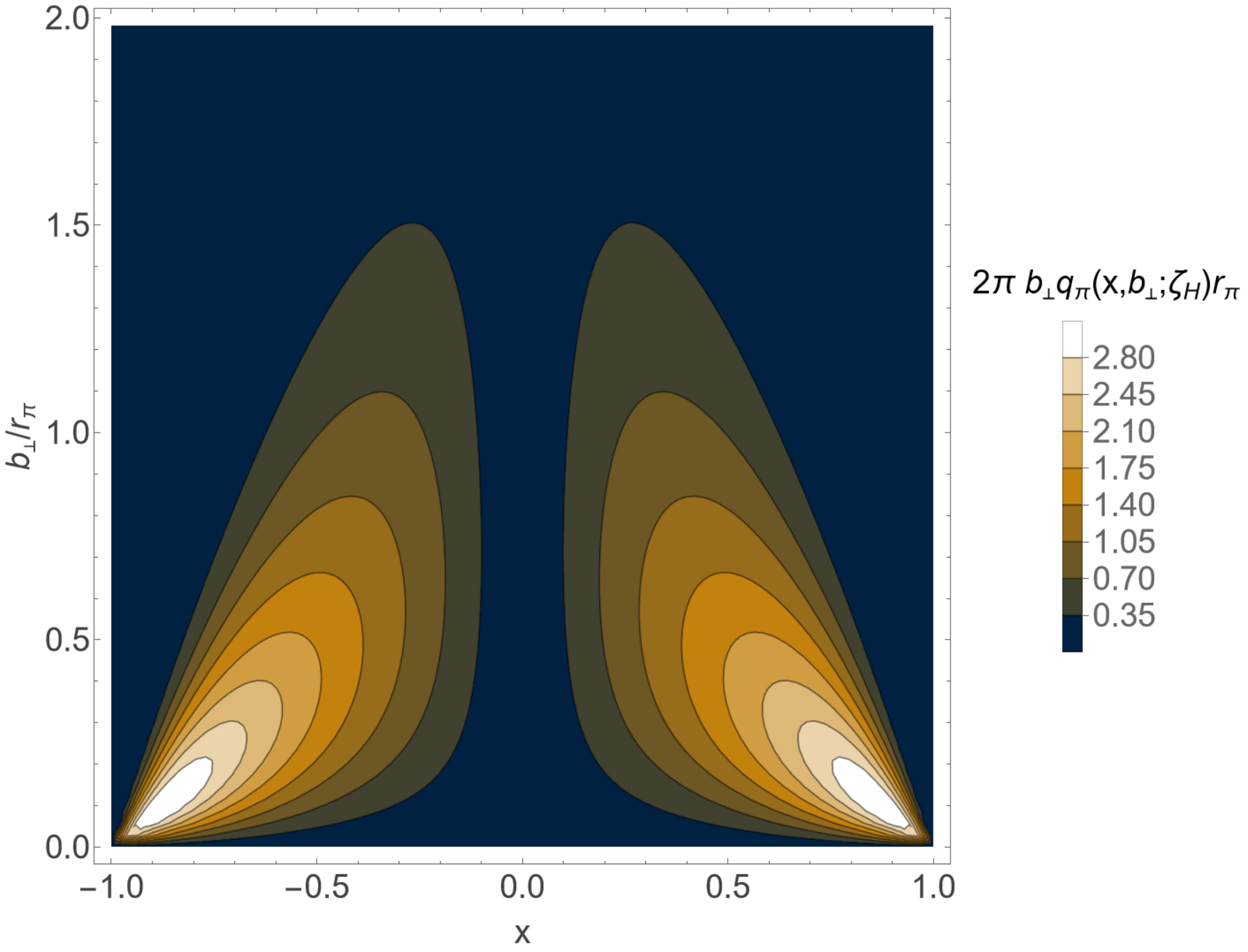} 
\end{tabular}
}
\caption{$\pi$ IPS-GPD obtained via Eq.\,\eqref{eq:IPSMA}. }
\label{fig:IPSGPDs} 
\end{figure}

The pion IPS-GPD shown in Fig.\,\ref{fig:IPSGPDs} exhibits notable features. For small $|x|$, the distribution is broad in $|b_\perp|$, indicating a low probability of finding valence constituents in this region at $\zeta_H$.  As $|x|$ increases, the distribution develops a clear maximum at some value of $|b_\perp|$. This peak rises in height and becomes narrower as $|x|$ increases. These patterns are typical of light systems. In systems with heavy quarks, the distributions are significantly more localized and exhibit higher peaks,\,\cite{Raya:2024glv}.

\subsection{Charge and mass distributions}
\begin{figure}[t]
\centerline{
\begin{tabular}{c}
\includegraphics[width=0.45\textwidth]{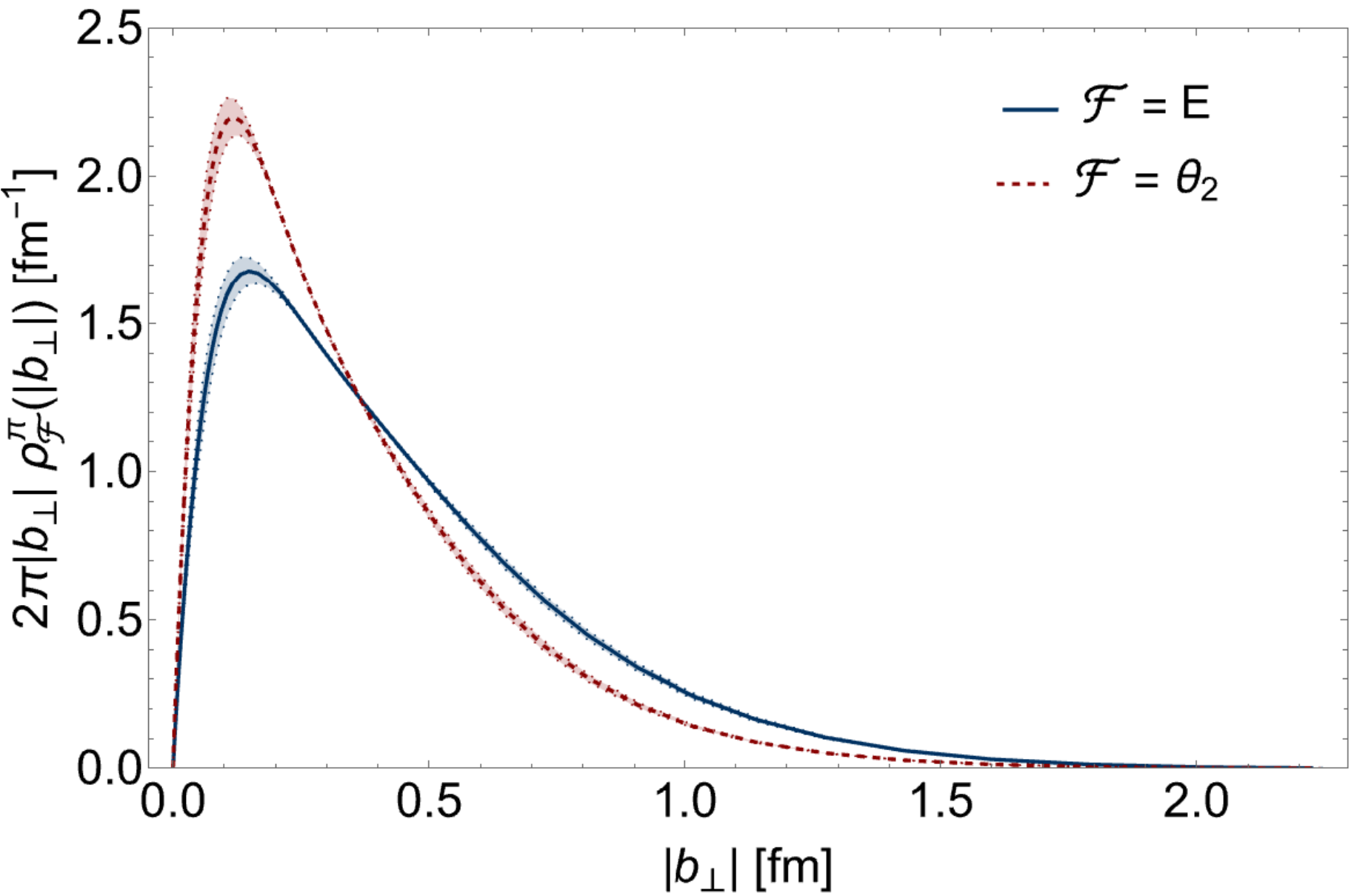}
\end{tabular}
}
\caption{$\pi$ charge and mass distributions as defined in Eq.\,\eqref{eq:chargeMass}.}
\label{fig:Dists} 
\end{figure}

The role of a valence-quark $q$ in the two-dimensional charge and mass distributions, $\rho_E^{\textbf{P}q}(b_\perp)$ and $\rho_{\theta_2}^{\textbf{P}q}(b_\perp)$, respectively, is encoded within the zeroth and first moments of the IPS-GPD\,\cite{Raya:2022eqa,Xu:2023bwv}; that is:
\begin{align}
\label{eq:chargeMass}
    \rho_{\{E,\theta_2\}}^{\textbf{P}q}(|b_\perp|)=&\int_0^{1}dx\, \{1,x\}\,q_{\textbf{P}}(x,|b_\perp|)\\
    =&\int_0^\infty\frac{d\Delta}{2\pi}\Delta J_0(|b_\perp|\Delta)\{F_\textbf{P}^q(\Delta^2),\,\theta_2^{\textbf{P}q}(\Delta^2)\}\,.\notag
\end{align}
As with the corresponding form factors, the meson's total distributions are:
\begin{align}\label{eq:chargeDist}
    \rho_{E}^{\textbf{P}}(|b_\perp|)=&\,e_q \rho_{E}^{\textbf{P}q}(|b_\perp|)+e_{\bar{h}}\rho_{E}^{\textbf{P}\bar{h}}(b_\perp)\,,\\
    \rho_{\theta_2}^{\textbf{P}}(|b_\perp|)=&\, \rho_{\theta_2}^{\textbf{P}q}(|b_\perp|)+\rho_{\theta_2}^{\textbf{P}\bar{h}}(|b_\perp|)\,.\label{eq:massDist}
\end{align}
The $\pi$-associated distributions are shown in Fig.\,\ref{fig:Dists}. The charge density extends slightly beyond the mass distribution, while the latter exhibits a more pronounced peak at small transverse separation. This pattern reflects a genuine separation between the spatial distributions of charge and mass, with the electromagnetic structure probing longer-distance dynamics than the more localized mass density. Nevertheless, as the meson mass increases, both distributions are expected to become narrower and more alike, ultimately approaching the point-particle limit,\cite{Raya:2024glv,Raya:2024ejx}.

Another measure of the relative compression of the mass with respect to the charge distribution is given by
\begin{equation}
 \frac{\rho_M^{\pi^+}(|b_\perp|)}{\rho_E^{\pi^+}(|b_\perp|)}\overset{|b_\perp|\to 0}{\approx}1.62(3)\,.
\end{equation}
This enhancement at small $|b_\perp|$ further supports the picture of a more localized mass distribution. As discussed in Ref.\,\cite{Raya:2024glv}, the corresponding partonic limit, where the non-perturbative dynamics of QCD is supressed, yields $3/2$.

\subsection{Transverse Momentum Dependent Distributions}
We now focus on the pion's leading-twist time-reserval even TMD, which in the present context is defined as\,\cite{Pasquini:2014ppa}:
\begin{equation}
\label{eq:TMDdef}
    f_\textbf{P}^q(x,k_\perp^2)=\frac{1}{16\pi^3}\left[|\psi_{\textbf{P}q}^{\uparrow\downarrow}(x,k_\perp^2)|^2+k_\perp^2|\psi_{\textbf{P}q}^{\uparrow\uparrow}(x,k_\perp^2)|\right]\,.
\end{equation}
Once again we have adopted $\zeta_H$ as the defining resolving scale. Given Eq.\,\eqref{eq:DFdef}, it is clear that the integration over $k_\perp$ recovers the DF, that is:
\begin{equation}
    q_\textbf{P}(x)=\int d^2k_\perp f_\textbf{P}^q(x,k_\perp^2)\,.
\end{equation}
These interrelations between the TMD, the LFWF, and the DF imply that the former inherits key features from the latter. Consequently, for fixed $k_\perp^2$, we find a large-$x$ behavior of $(1-x)^2$. Likewise, since $\psi_{\textbf{P}q}^{\uparrow\downarrow}(x,k_\perp^2)$ falls as $1/k_\perp^2$ and $\psi_{\textbf{P}q}^{\uparrow\uparrow}(x,k_\perp^2)$ as $1/k_\perp^4$, the large-$k_\perp^2$ falloff of the TMD is $f_\textbf{P}^q(x,k_\perp^2)\sim1/k_\perp^4$. These patterns are consistent with QCD prescriptions\,\cite{Brodsky:2006hj,Lepage:1979zb,Lepage:1980fj,Soper:1976jc,Farrar:1975yb,Berger:1979du}. 

\begin{figure}[t]
\centerline{
\begin{tabular}{c}
\includegraphics[width=0.45\textwidth]{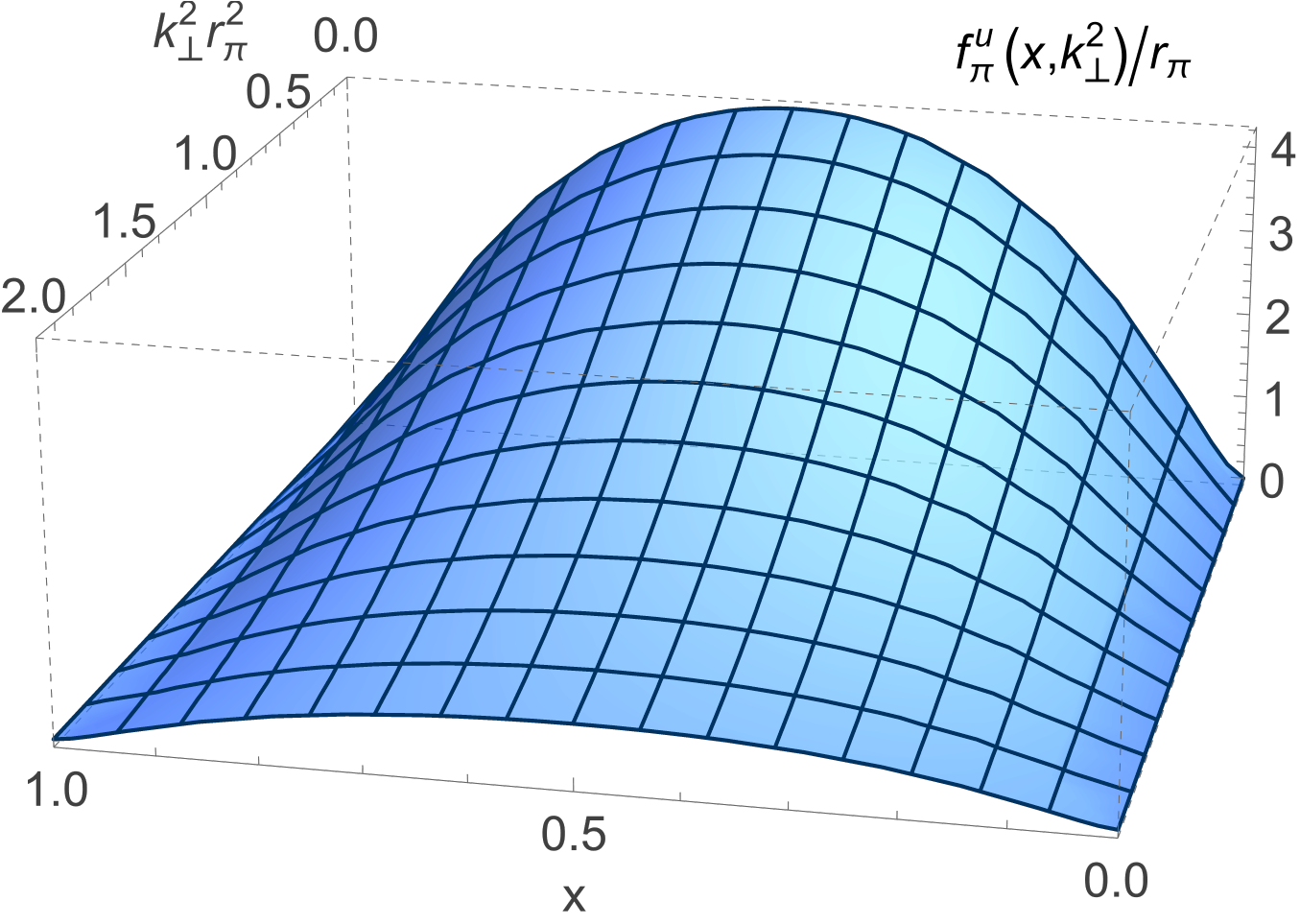} 
\end{tabular}
}
\caption{$\pi$ time reversal even TMD as defined in Eq.\,\eqref{eq:TMDdef}.}
\label{fig:TMD} 
\end{figure}

The profile of the pion TMDs is shown in Fig.\,\ref{fig:TMD}. Clearly, for every fixed $k_\perp^2$, the characteristics of the corresponding DF are plainly reflected: a broad distribution in the intermediate-$x$ region, and a smooth falloff at the endpoints. The monotonic decrease with increasing $k_\perp^2$ is also evident. 

The pion TMD shown in Fig.\,\ref{fig:TMD} displays the expected structure in both $x$ and $k_\perp$. For fixed $k_\perp$, the TMD is broad in the intermediate-x region and exhibits a smooth falloff at the endpoints. Moreover, the overall magnitude decreases systematically with increasing $k_\perp$.

Defining the expectation value $\langle k_\perp^m\rangle_{\textbf{P}}^q$ as:
\begin{equation}
    \label{eq:defkperp}
    \langle k_\perp^m\rangle_{\textbf{P}}^q=\int_0^1dx\int d^2k_\perp f_\textbf{P}^q(x,k_\perp^2)(k_\perp^2)^{m/2}\,,
\end{equation}
we find $\langle k_\perp^1\rangle_\pi^{q}=0.254 \,\text{GeV}$ and $\langle k_\perp^2\rangle_\pi^{q}=(0.292\,\text{GeV})^2$. The present framework produces values that are somewhat smaller than CSMs explorations, see e.g.\,\cite{Ding:2025zcc}. In our case, however, their magnitudes are naturally related to the mass scale that governs the pion LFWFs; namely, $M_q \sim 0.3$ GeV. Furthermore, we find the interesting connection $\langle k_\perp^2\rangle_\pi^{q} \approx (r_E^\pi)^{-2}$. 

Note that the resolving scale $\zeta_H$ is essentially a definition and need not be assigned a specific value. Nevertheless, phenomenological analyses typically find $\zeta_H \sim 0.35$ GeV (see \,\cite{Lu:2023yna} and references therein). Under TMD evolution, the transverse distribution becomes broader and less peaked as the scale increases~\cite{Anselmino:2005nn}, resulting in larger $\langle k_\perp^2\rangle_\pi^{q}$ values. For instance, phenomenological extractions at higher scales $\zeta^2>2\,\text{GeV}^2$ yield: $\langle k_\perp^2\rangle_\pi^q = (0.5\,\text{GeV})^2$\,\cite{Anselmino:2005nn}, $(0.57\,\text{GeV})^2$\,\cite{Collins:2005ie}, and $(0.62\pm0.05\,\text{GeV})^2$\,\cite{Schweitzer:2010tt}. The scale evolution of the TMD, and its consequences, shall be explored in a future work.

\section{Conclusions and Scope}
\label{sec:conclusions}
This work describes the construction of the leading-twist LFWFs of pseudoscalar mesons. Both helicity-0 and helicity-1 components are considered, and all structures characterizing the Bethe-Salpeter amplitude are effectively incorporated. In doing so, we show that while the helicity-0 LFWF and the dominant BSA are often sufficient to capture the bulk of the phenomenology, all components are required in order to reproduce the expected asymptotic behaviors.

The construction of the LFWFs within the PTIR framework indicates that the $x$–$k_\perp$ correlations are driven by explicit chiral symmetry breaking and, therefore, vanish in the chiral limit. In such scenario, the PTIR-based LFWFs reduce to a simple form in which both the asymptotic limits and the factorization of the $x$ and $k_\perp$ dependence become manifest. Building on this outcome, we then introduce the MA, in which correlations between these kinematic variables are effectively encoded in a characteristic function $\mathcal{M}_q(x)$. The MA framework also allows one to directly employ existing information on either the DA or DF in the construction of the corresponding LFWFs. Furthermore, it enables the derivation of closed expressions and algebraic relations among different quantities.

Focusing on the pion, we then analyze the DGLAP GPD and the associated electromagnetic and gravitational form factors. These are subsequently mapped to impact-parameter space, yielding the corresponding IPS-GPDs as well as charge and mass distributions. In all cases, the expected patterns are reproduced. Among these are the broadness of the pion distributions --a pattern that emerges due to its NG character-- the agreement with empirical information on the pion EFF, and the larger spatial extent of the charge density relative to the mass distribution. Thus, given suitable input for the DA or DF, both the PTIR and MA reproduce the results of  computationally intensive methods. We also briefly discuss the pion TMD, showing that the magnitude of its dominant moments at the hadronic scale is closely tied to the mass scales that define the LFWFs. Further explorations beyond $\zeta_H$ are expected.

The pion-related quantities discussed here are expected to be within reach of modern experimental facilities, in particular the new generation of Electron–Ion colliders. Their study would constitute a significant step toward elucidating the complexities of strong interactions, especially their emergent features associated with mass generation and confinement.

\medskip
\noindent\textbf{Acknowledgments}.
Work supported by the Spanish Ministry of Science and Innovation (MICINN grant no.\ PID2022-140440NB-C22), and Junta de Andalucía (grant no.\ P18-FR-5057). Z.Q. Yao acknowledges support from Helmholtz-Zentrum Dresden-Rossendorf, under the High Potential Programme.

\medskip
\noindent\textbf{Data Availability Statement}. This manuscript has no associated data or the data will not be deposited. All information necessary to reproduce the results described herein is contained in the material presented above.

\medskip
\noindent\textbf{Declaration of Competing Interest}.
The authors declare that they have no known competing financial interests or personal relationships that could have appeared to influence the work reported in this paper.

\bibliographystyle{unsrt}
\bibliography{pion.bib}

\end{document}